\documentclass[%
 superscriptaddress,
 preprint,
 amsmath,amssymb,
 aps,
 prl.
]{revtex4-2}

\usepackage{graphicx}
\usepackage{dcolumn}
\usepackage{bm}
\usepackage{siunitx}        
\usepackage{color}          
\usepackage{babel}
\usepackage{hyperref}

\begin{document}


\title{Longitudinal magneto-thermal conductivity and magneto-Seebeck of itinerant antiferromagnetic BaMn$_2$Bi$_2$}

\author{Takuma Ogasawara}
\affiliation{Beijing Academy of Quantum Information Sciences, Beijing 100193, China}

\author{Hailiang Xia}
\affiliation{Institute of Physics Chinese Academy of Sciences, China}

\author{Khuong-Kim Huynh}
\affiliation{Department of Chemistry, Aarhus University, DK-8000 Aarhus, Denmark}

\author{Qifeng Yao}
\affiliation{Beijing Academy of Quantum Information Sciences, Beijing 100193, China}

\author{Liguo Zhang}
\affiliation{Beijing Academy of Quantum Information Sciences, Beijing 100193, China}

\author{Thomas L M Lane}
\affiliation{Beijing Academy of Quantum Information Sciences, Beijing 100193, China}

\author{Shilin Li}
\affiliation{Beijing Academy of Quantum Information Sciences, Beijing 100193, China}
\affiliation{Beijing National Laboratory for Condensed Matter Physics, Institute of Physics ( IOP), Chinese Academy of Sciences, 603, Beijing 100190, China}
\affiliation{University of Chinese Academy of Sciences  (UCAS), Beijing 100049, China}

\author{Yufeng Gao }
\affiliation{Beijing Academy of Quantum Information Sciences, Beijing 100193, China}
\affiliation{Beijing National Laboratory for Condensed Matter Physics, Institute of Physics ( IOP), Chinese Academy of Sciences, 603, Beijing 100190, China}
\affiliation{University of Chinese Academy of Sciences  (UCAS), Beijing 100049, China}

\author{Tingting Hao}
\affiliation{Beijing Academy of Quantum Information Sciences, Beijing 100193, China}

\author{Jianhao Chen}
\affiliation{Beijing Academy of Quantum Information Sciences, Beijing 100193, China}
\affiliation{International Center for Quantum Materials, School of Physics, Peking University, Beijing 100871, China}
\affiliation{Key Laboratory for the Physics and Chemistry of Nanodevices, Peking University, Beijing 100871, China}

\author{Katsumi Tanigaki}
\email{katsumitanigaki@baqis.ac.cn} 
\affiliation{Beijing Academy of Quantum Information Sciences, Beijing 100193, China}


\date{\today}

\begin{abstract}
Thermal transport, generally mediated by the direct microscopic exchange of kinetic energy via lattice phonons, can also be modified by contributions from additional quasiparticles, such as electrons and magnons.
However, a comprehensive understanding of the magnon influence has yet to be realized and remains an active research area.
The most significant roadblock has been a lack of available materials in which these three quasiparticles can be clearly identified and quantitatively examined in order to provide an intrinsic understanding, not only of their independent contributions to thermal conductivity but also of the cross-correlated interactions among them.
Itinerant antiferromagnetic (AFM) BaMn$_{2}$Bi$_{2}$ with PT symmetry exhibits Anderson metal-insulator localization, which can be tuned into the metallic regime via an applied magnetic field due to its unique electron-magnon interactions. 
We identify itinerant AFM BaMn$_{2}$Bi$_{2}$ as an ideal material for scientific investigations into how these quasiparticles participate in thermal conductivity.
Here, we present the direct contribution of electrons, phonons, and magnons to thermal conductivity, as well as their interspecies interactions, supported by detailed analyses conducted in the framework of the Boltzmann transport formalism.
The comparison of the magneto-thermal conductivity and magneto-electrical conductivity, as well as the magneto-Seebeck effect of itinerant antiferromagnetic BaMn$_{2}$Bi$_{2}$, gives unique insight into how magnons participate in longitudinal thermal-associated phenomena.

\end{abstract}

\maketitle


\section*{\label{sec:level1}Introduction}
Thermal conductivity $\kappa$, one of the fundamental transport properties, quantifies the flow of energy through a medium due to a temperature gradient, $\nabla T$.
This flow is generally mediated by phonons, the quasiparticle (QP) associated with excitations of lattice vibrations, and by the kinetic motion of electrons and holes located near the Fermi surface.
However,  $\kappa$ can also be realized via other QPs such as magnons, which are associated with the transport of spin-angular momentum and are of significant importance for the transverse thermal-Hall effect and other thermoelectric-related phenomena~\cite{zhang_berry_2016}.
To date, spin angular-momentum dynamics~\cite{zutic_spintronics_2004}, such as spin-caloritronics~\cite{bauer_spin_2012}, have been investigated extensively, opening new avenues towards the advanced quantum information science and technology.
The dynamics of these various QPs play a significant role in determining the thermal properties of materials, however quantifying their impact still poses a challenge since $\kappa$ is determined not only by the direct contribution from the phonons, electrons and magnons (given by $\kappa_\mathrm{ph}$, $\kappa_\mathrm{el}$ and $\kappa_\mathrm{mag}$ respectively), but is also impacted by the interspecies interactions amongst them.
Understanding the thermal transport phenomena of these QPs remains a key research field, being important not only for fundamental science but also holding technological significance for a broad field of applications related to highly efficient energy conversion.

Amongst these three QPs, magnons are possibly the most important due to their ability to break time-reversal symmetry.
Magnons influence $\kappa$ in two key ways: via a direct contribution from magnon transport, and through a modification of the relaxation time $\tau$ due to ph-mag and/or el-mag interactions.
Likewise, magnons also impact both the electrical conductivity, $\sigma$, and the Seebeck coefficient, $S$, of longitudinal thermoelectric charge-transport.
For example, significantly large magnetoresistance as a consequence of el-mag scatterings tuned by a magnetic field, $H$, is an important research theme in 3$d$ ferromagnets (FM)~\cite{raquet_electron-magnon_2002}.
Notably, Yttrium Iron Garnet (YIG), one of the most important insulator-based FM materials for spintronics, exhibits an appreciable reduction in $\kappa$ in the presence of non-zero $H$~\cite{ratkovski_thermal_2020}.
Theoretical investigations of magnons and their relation to other transport phenomena are also an active area of research~\cite{rezende_thermal_2015,wu_magnon_2018,wang_exchange_2010,pan_ab_2023,sun_generic_2021}.
Anti-ferromagnetic (AFM) materials offer various advantages when compared to FM materials, such as robustness against $H$-fields and an ultrafast switching speed~\cite{olejnik_terahertz_2018}.
In addition, as a consequence of their zero total magnetic moment the manipulation of the magnetic moments in AFM materials is distinct from that of FM materials~\cite{wadley_electrical_2016}.

$\mathrm{BaMn_2Bi_2}$ (BMB) exhibits intriguing orbital-selective anti-ferromagnetism in a 3$d$-Mn system, with a $d_{xy}$ hole band influenced by el-mag interactions.
The remaining $d$-orbitals show strong electron correlations, resulting in a Mott-insulator with AFM spin order~\cite{ogasawara_magnetic_2022}.
Holes in the $d_{xy}$-derived band exhibit a metallic state with small electrical conductivity at high temperatures, where chemical potential $\mu$ is naively below the top of the $d_{xy}$ hole band due to the carrier accumulation by lattice defects.
Most intriguingly, the metal-insulator transition (MIT) taking place due to the Anderson localization of the system is tunable by an external magnetic field ($H$) applied perpendicular to the easy axis of magnetized G-type AFM moments, $H_{ab}$, leading to the material becoming metallic under high field strengths.
By employing scaling analysis for $\kappa$ at different $H_{ab}$, the critical exponent of the MIT has been evaluated to be $\nu\approx1.4$, which can be classified as the first example of $H$-associated Anderson localization in a magnetically disordered system~\cite{wang_universality_2021}.
The unusual tuning induced by $H_{ab}$ is indicative of a strong coupling between the AFM spin-angular momentum fluctuations and the itinerant $d_{xy}$ holes in this $d$ multi-orbital system.
As a result, BMB respresents a unique scientific platform to study how electrons, phonons, and magnons, as well as their inter-species interactions, can participate in thermal conductivity in the longitudinal direction.

Here, we report the longitudinal thermal-associated transport of itinerant anti-ferromagnetic BMB, in both the absence and presence of an in-plane magnetic field, $H_{ab}$, applied perpendicular to the AFM easy axis.
Through an experimental investigation of the $T$-evolution of the thermal conductivity ($\kappa$), electrical conductivity ($\sigma$), and Seebeck coefficient ($S$), in conjunction with a theoretical model of the electron, phonon, and magnon transport, we provide an interpretation of how magnons contribute to $\kappa$.

\section*{\label{sec:level2}Results}
\begin{figure}[ht]
	\centering
	\includegraphics[width=0.48\textwidth]{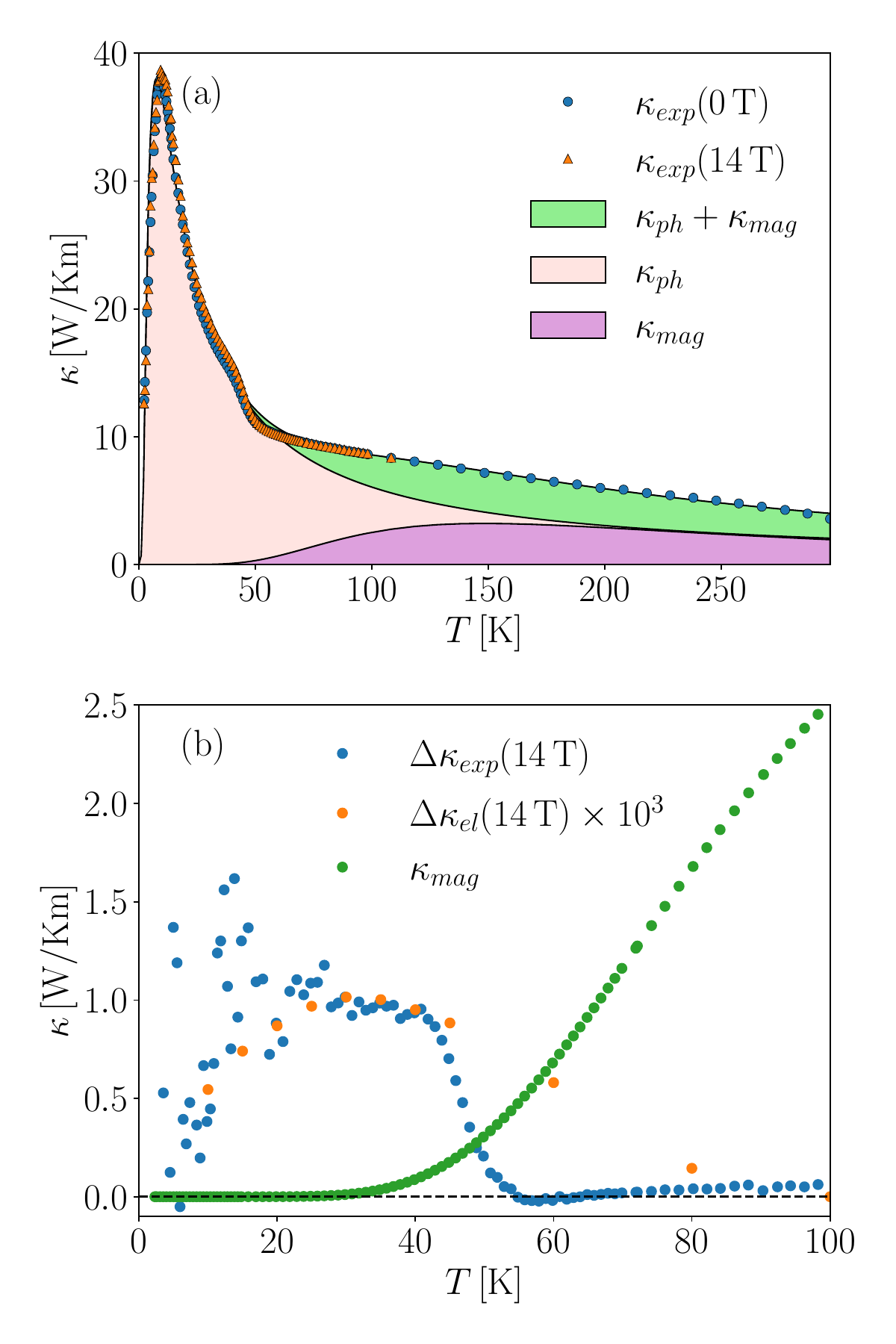}
	\caption{(a) Phonon and magnon thermal conductivity compared with experimental results, measured with and without magnetic field along the $a$-axis.\ (b) Size comparison of the field induced values of thermal conductivity in experiment $\Delta \kappa_\mathrm{exp}=\kappa_\mathrm{exp}(14T)-\kappa_\mathrm{exp}(0T)$, the rescaled calculated electron contribution $\Delta \kappa_\mathrm{el}=\kappa_\mathrm{el}(14T)-\kappa_\mathrm{el}(0T)$ and the zero-field magnon contribution.}\label{fig:kappa}
\end{figure}

The $\kappa$ of itinerant AFM-BMB was measured under applied field strengths of $\mu_0H=0$ and $\mu_0H=\SI{14}{T}$, aligned parallel to the crystal's $ab$-plane (perpendicular to the crystallographic $c$-axis).
In the former case, $\sigma$ exhibits Anderson localization, whilst for $\mu_0H\geq~\SI{0}{T}$ the system returns back to a metallic state~\cite{ogasawara_magnetic_2022}.
By contrast, we note that when $H$ is oriented perpendicular to the $ab$-plane, we observed no modification to the transport properties ~\cite{wang_thermopower_2013}.
In Fig.~\ref{fig:kappa}(a) we show the measured experimental value of $\kappa_\mathrm{exp}$ along with the results of theoretical simulations, the details of which shall follow later.
The profile of $\kappa_\mathrm{exp}$ in the presence of non-zero $H$ deviates from the zero-field case most significantly for temperatures in the range $30\mathrm{K}<T<40\mathrm{K}$.
This is illustrated in Fig.~\ref{fig:kappa}(b), where we plot the value of $\Delta \kappa_\mathrm{exp}=\kappa_\mathrm{exp}(14T)-\kappa_\mathrm{exp}(0T)$.

In order to evaluate $\kappa_\mathrm{qp}$, we performed theoretical calculations based upon the band dispersions depicted in Fig.~\ref{fig:band} (see further details in the Supplemental Information).
The electron contribution to magneto-thermal conductivity, $\kappa_\mathrm{el}$, at $\mu_0H=\SI{14}{T}$ was calculated by employing the band structure presented in~\cite{ogasawara_magnetic_2022}.
We find that $\kappa_\mathrm{el}$ is $3$ orders of magnitude smaller across the entire $T$ range when compared to the measured values (Fig.~\ref{fig:kappa}~(b)).
Therefore, the sum of the magnon and phonon contributions alone demonstrate a good approximation to the experimental results, i.e. $\kappa_\mathrm{exp}\approx\kappa_\mathrm{ph}+\kappa_\mathrm{mag}$.
These results are also consistent with our analysis of the heat capacity, $C_{p}$ (further details to be found in the Supplemental Information).

\begin{figure*}[ht]
	\centering
	\begin{minipage}[h]{0.3\textwidth}
	\includegraphics[width=0.98\textwidth]{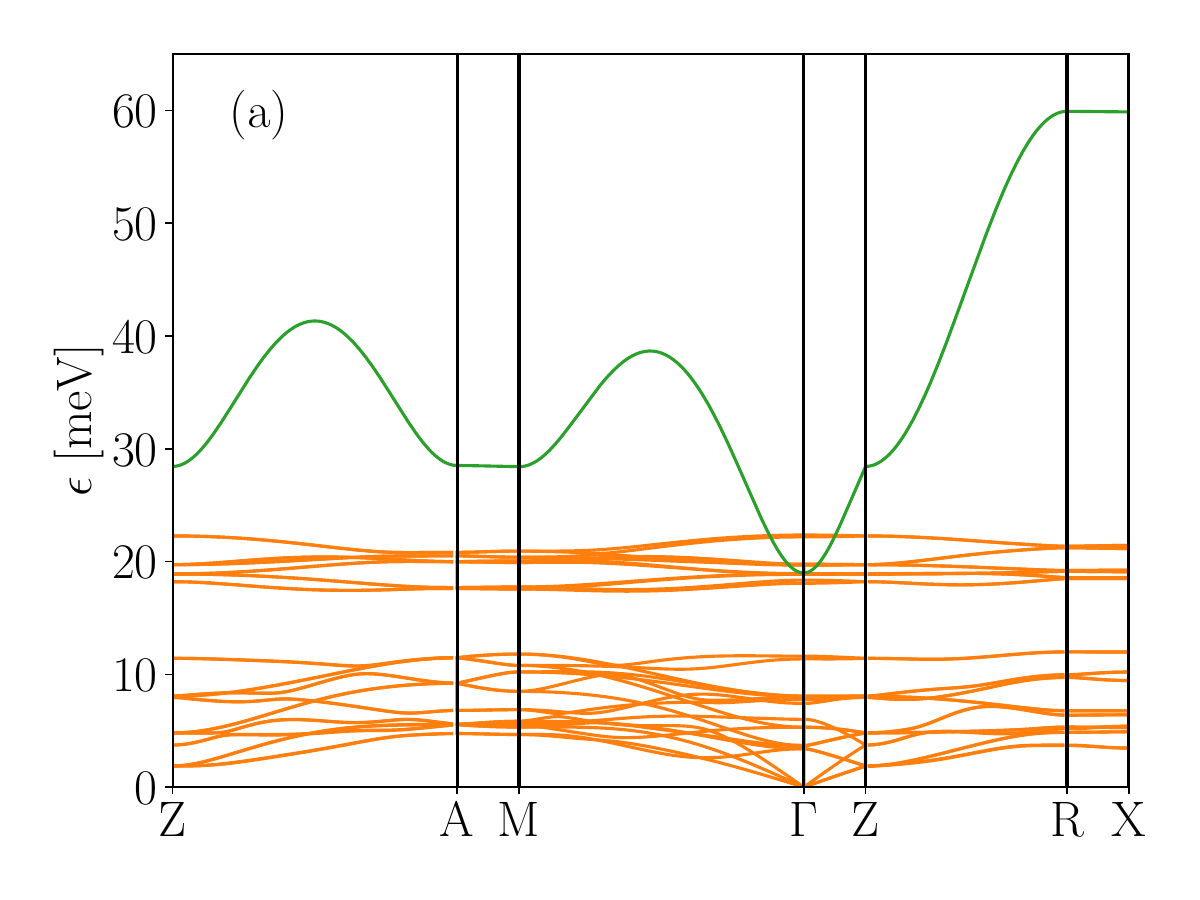}
	\end{minipage}
	\begin{minipage}[h]{0.3\textwidth}
	\includegraphics[width=0.98\textwidth]{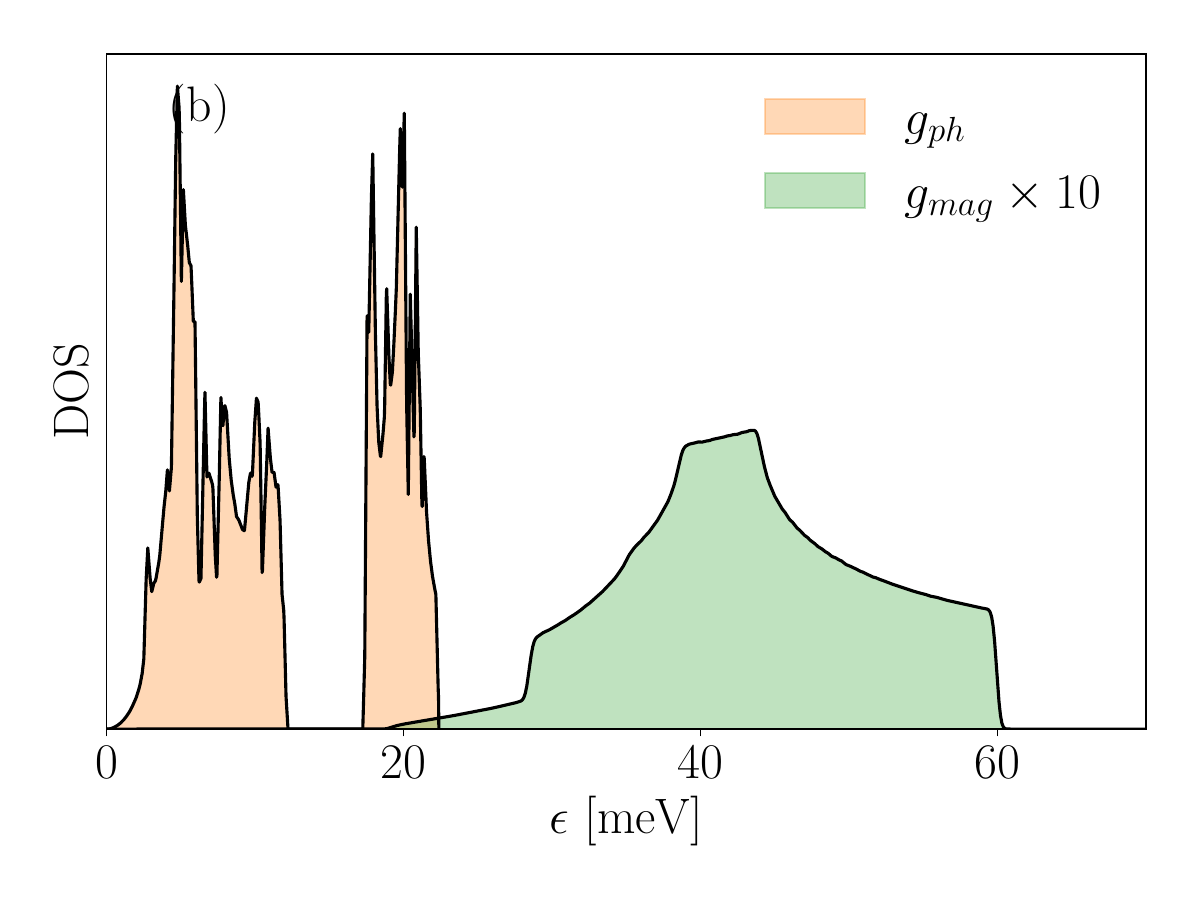}
	\end{minipage}
	\begin{minipage}[h]{0.3\textwidth}
	\includegraphics[width=0.98\textwidth]{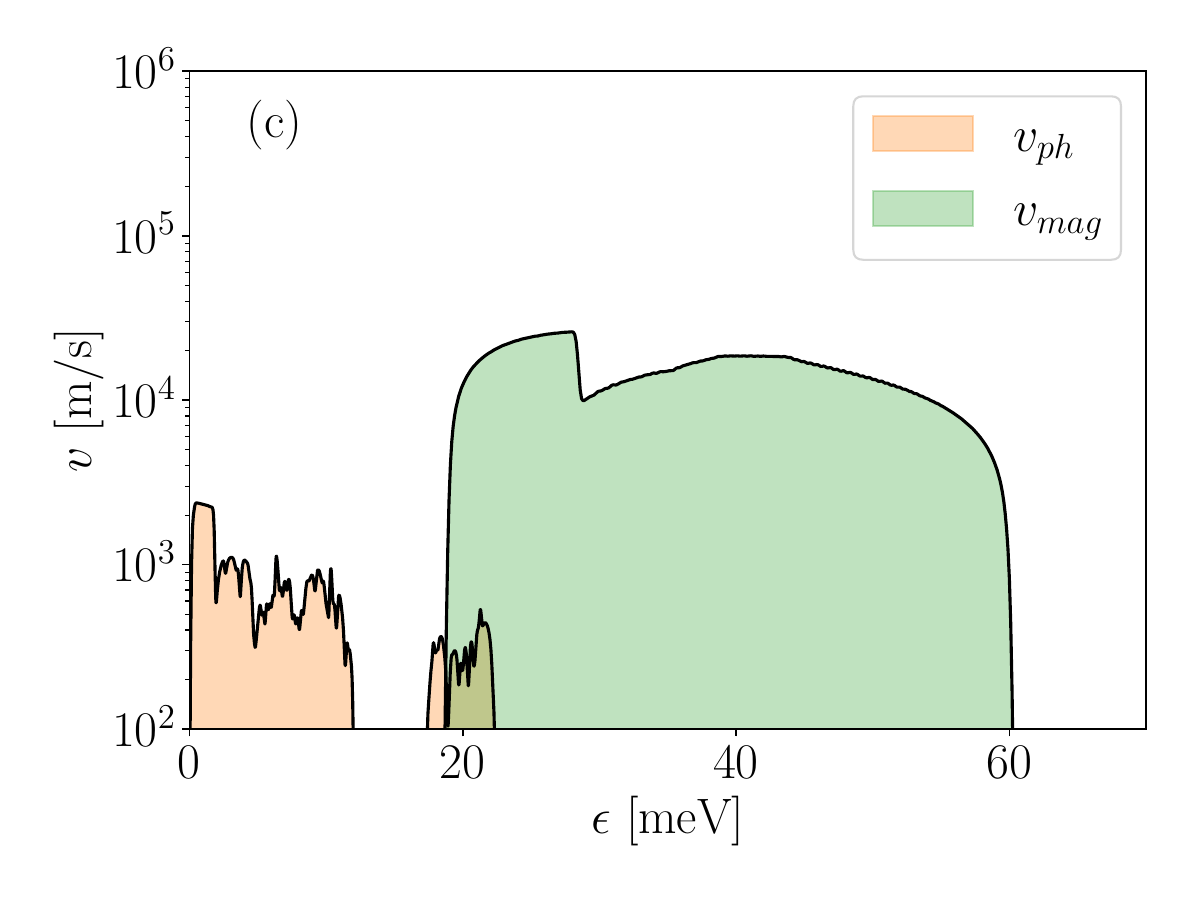}
	\end{minipage}
	\caption{(a) The phonon and magnon dispersion relations in $BaMn_{2}Bi_{2}$. (b) Density of states ($g$) of the electrons (el), phonons (ph), and magnons (mag). (c) Average of group velocity $v$.}\label{fig:band}
\end{figure*}

In the low-$T$ regime, $\kappa$ is dominated by the phonon contribution.
It exhibits a maximum at $T\sim\SI{10}{K}$, where the dominant scattering process switches to intrinsic phonon scatterings.
Although a similar peak may, in principle, be argued in the magnon profile, its contribution is much smaller than that of phonons at low $T$ based on our theoretical calculations.
This is because magnon excitations are suppressed at low $T$s due to the $\SI{18}{meV}$ gap observed experimentally in the magnon dispersion.
As such, the parameters were optimized by focusing principally on fitting $\kappa_\mathrm{ph}$ to the experimental data in the low-$T$ regime.
While $\kappa_\mathrm{ph}$ alone is sufficient to reproduce the low-$T$  profile, it diverges significantly at higher $T$s.
We therefore optimized the parameters for $\kappa_\mathrm{mag}$ by fitting to $\kappa_\mathrm{exp}-\kappa_\mathrm{ph}$ in this regime.
The combination of $\kappa_\mathrm{ph}$ and $\kappa_\mathrm{mag}$ well reproduces $\kappa_\mathrm{exp}$ for $\mu_0H=\SI{0}{T}$.
As shown in Fig.~\ref{fig:kappa}~(a), whilst $\kappa_\mathrm{ph}$ and $\kappa_\mathrm{mag}$ each individually exhibit a peak, they result in only a single maxima in $\kappa_\mathrm{exp}$ since the magnon peak is significantly suppressed behind the dominant $\kappa_\mathrm{ph}$ profile.
This is not generally the case for other AFM insulators such as La$_2$CuO$_4$, which commonly exhibits a more prominent double peak profile~\cite{hess_magnon_2003}.

When a field of $\mu_0H=\SI{14}{T}$ is applied parallel to the $ab$-plane of BMB, an enhancement to the magneto-thermal conductivity appears below $T\sim\SI{50}{K}$.
We note once again that $\kappa$ did not change when $H$ is parallel to the $c$-axis.
To examine the magnetic field effects on magnon, the magnon dispersion under magnetic field was calculated by referring to the previous method~\cite{rezende_introduction_2019} (see the details in the Supplemental Information).
The calculations show that the magnon scattering arising from $\bm{k}\approx0$ under $\mu_0H=\SI{0}{T}$ at the two energy crossing points is lifted when $\mu_0H=\SI{14}{T}$, leading to an increase in $\kappa$.

In the case of YIG, negative magneto-thermal conductivity was reported below $T\sim\SI{20}{K}$ due to the $H$ field-induced opening of a gap in the magnon dispersion, leading to a reduction in the number of magnons participating in thermal conduction~\cite{ratkovski_thermal_2020}.
For BMB, given the experimental evidence that the magnon contribution is negligibly small below $T\sim\SI{50}{K}$, it is more plausible that the $H$ dependence can be interpreted as originating from modulation due to mag-ph interactions or due to additional magnons excited by the $H$ field.

\begin{figure}[ht]
	\centering
	\includegraphics[width=0.48\textwidth]{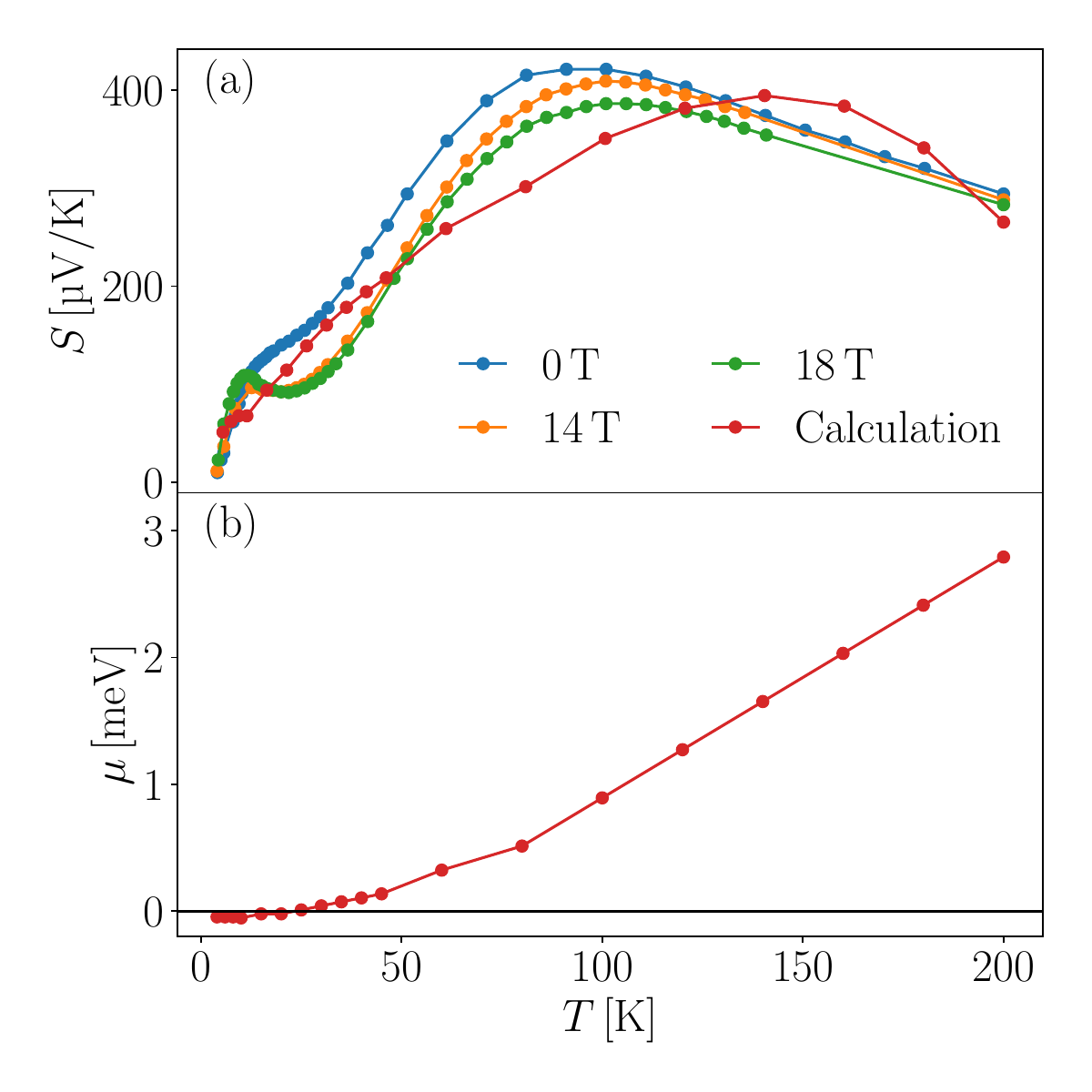}
	\caption{(a) Seebeck coefficient, $S$, of BMB with and without magnetic field along the $a$-axis.\ (b) The estimated chemical potential, $\mu$, used to calculate $S$. We choose the top of valence band as the energy origin.}\label{fig:seebeck}
\end{figure}

In Fig.~\ref{fig:seebeck} we show the temperature dependence of the magneto-Seebeck coefficient $S(H)$ at different field strengths, from which we can more clearly identify the effect of the el-mag interactions.
Notably, the negative magneto-Seebeck effect reaches approximately $-38\%$ under an $\mu_0 H=\SI{18}{T}$.
Theoretical calculations conducted using BoltzTraP2 in the framework of the BTE qualitatively reproduce the experimental results of $S(H)$ (the details of which can be seen in the Supplemental Information).
Since the chemical potential is always located within the valence band, $S(H)$ is positive irrespective of $T$.
It does, however, typically lie close to the valence band edge and therefore the magnitude of $S(H)$ is very sensitive to fluctuations in $T$.
$S(H)$ takes its maximum value around $\SI{100}{K}$ (representing a small deviation from the calculated value of $\SI{140}{K}$), where the edge of the window function $(\epsilon -\mu)\left(-\frac{\partial f}{\partial \epsilon}\right)$ in $K_n$ reaches the conduction band edge.
This leads to a small decrease in $S(H)$ by balancing the contributions from the hole and electron bands.
As stated in Supplemental Information, transport properties of electrons are described as $\sigma = e^2K_0$ and $S = -\frac{1}{eT}\frac{K_1}{K_0}$ using kinetic coefficient $K_n$.
According to $K_n$, both  $K_0$ and $K_1$ exhibit a $T$-dependence, however the latter retains a more significant effect at higher temperatures.
Consequently, the magneto-$S$ effect is present at a higher T than that of the magneto-$\sigma$.
The experimental dependence upon $H$, observed both for $\kappa(H,T)$ and $S(H,T)$, suggests that the magnetic field alters the electronic states contributing to the electrical transport phenomena without any significant influence upon other excitations, such as phonon or magnon.
This is consistent with the orbital-selectivity of electronic states in this compound~\cite{ogasawara_magnetic_2022}.

\section*{\label{sec:level2}Discussion}
The characteristic distances defining the lattice and magnetic boundaries, fitted from the experimental data, are $L_l=1.54\times 10^{-2}\si{mm}$ and $L_m=502\text{\AA}$, respectively.
The former is an order of magnitude smaller than the size of the sample used in this investigation and the latter is the same order as the AFM domain size reported for other AFM insulators~\cite{fitzsimmons_afm_2008}.
The fitting parameter employed for ph-ph scattering was $a=1.18\times10^{-17}\si{sK^{-1}}$ (see Eq.~\eqref{eq:ph-ph} of the Supplemental Information), which is one order of magnitude larger than in the case of archetypal ionic crystal, KCl, which had been studied previously~\cite{walker_phonon_1963}.
It indicates that an anharmonic potential of the ions in this system is relatively important, resulting in an enhanced ph-ph scattering rate.
In a similar fashion, the fitting parameter for mag-mag scattering, $c=7.22\times10^{8}\si{s^{-1}K^{-2}}$ (see Eq.~\eqref{eq:mag-mag} of the Supplemental Information), is two orders of magnitude larger than for that of YIG~\cite{rezende_thermal_2015}, indicating an increased mag-mag scattering rate.

\begin{figure}[ht]
	\centering
	\includegraphics[width=0.48\textwidth]{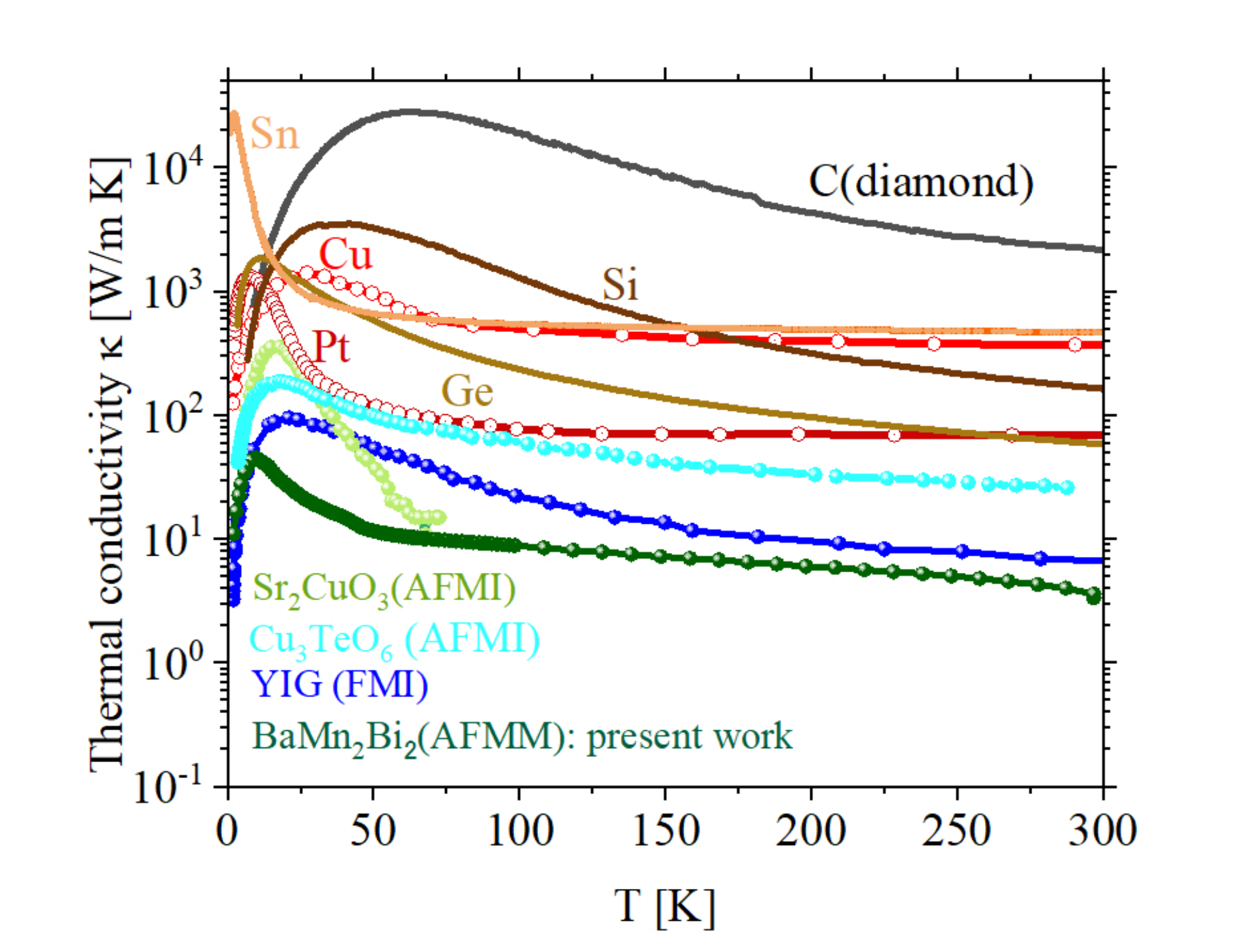}
	\caption{Thermal conductivity $\kappa$ of various materials as a function of $T$. They are classified into three types of materials: The first group contains the $IV_{th}$-group element semiconductors of Si, Ge, Sn (data are taken from~\cite{Elements}) in addition to insulating diamond of C (data are taken from~\cite{Diamond}). The second group consists of the conventional metals Cu and Pt (data are taken from~\cite{Elements}). The third group contains antiferromangetic insulators (data are taken from~\cite{AFM-CuTeO,YIG,Sr2CuO3}). Itinerant antiferromagnetic BMB, which is the focus of the present research, is the intriguing case and three quasiparticles of phonons, electrons, and magnons participates significantly in the longitudinal transport phenomena.}\label{fig:k-materials}
\end{figure}

Let us now discuss the above applies to the design and development of high-quality thermal conductors based on the framework of electron, phonon and magnon transport properties.
In Fig.~\ref{fig:k-materials} we compare the evolution of $\kappa$ for various materials as a function of $T$.
In the framework of the BTE described here, $\kappa$ consists of a sum over the independent quasiparticle contributions.
These contributions are determined primarily by the $g_\mathrm{qp}$, $v_\mathrm{qp}$, and $\tau_\mathrm{qp}$ of the quasiparticles qp $\in$ (el, ph, mag), and are critical for determining the total thermal conductivity, $\kappa_{\mathrm{total}}$.

Generally speaking, $g_\mathrm{mag}$ (the density of states of mag) $<$ $g_\mathrm{ph}$ (the density of states of ph) due to the disparity in the number of branches, which is equal to the number of localised spins for magnons and equal to the number of atoms multiplied by the vibrational degrees of freedom for phonons.
The scale of the dispersion is also important, since if the dispersion is large then $g_\mathrm{qp}$ spans a wider energy range.
In the case of BMB, the number of magnon and phonon branches are $2$ and $15$, respectively, with both dispersions lying within a $10\si{meV}$ energy window. 
It therefore follows that $g_\mathrm{ph}$ dominates over $g_\mathrm{mag}$ for this material.
As for electrons, $g_\mathrm{el}$ strongly depends upon the chemical potential $\mu$.
Since the $\mu$ of BMB is located in the vicinity of the valence band maximum, $g_\mathrm{el}$ is much smaller than both $g_\mathrm{ph}$ and $g_\mathrm{mag}$.
In a similar fashion, $v_\mathrm{qp}$ depends on the energy scale of the quasiparticle dispersion, $\epsilon_\mathrm{qp}$.
Qualitatively, the average $v_\mathrm{qp}$ for a band can be evaluated using $\left<v_\mathrm{qp}\right> \approx \frac{1}{\hbar}\frac{W}{k_\mathrm{max}}$, where $W = \epsilon_\mathrm{max}-\epsilon_\mathrm{min}$ and $k_\mathrm{max}$ are the bandwidth and distance from the BZ center to the boundary, respectively.
For instance, if we take the lattice constant to be of the order $a=\alpha\text{\AA}$ (with constant $\alpha$ generally lying in the range $10^0\sim10^1$), then since $\Delta k \approx \pi/a$ we have,
$    \left<v_\mathrm{qp}\right> \approx \frac{2\pi}{h}\frac{\alpha}{\pi}w \approx \frac{\alpha w}{2.07}\times 10^5\si{ms^{-1}eV^{-1}}$.

In the case of BMB, the electron, phonon and magnon bandwidth at the lowest energy level are approximately $W_\mathrm{el}\approx1\si{eV}$, $W_\mathrm{ph}\approx5\si{meV}$, and $W_\mathrm{mag}\approx40\si{meV}$, respectively.
Taking the lattice constant to be $a=4.49\text{\AA}$, we find $\left<v_\mathrm{el}\right>\approx2.17\times 10^5 \si{m/s}$, $\left<v_\mathrm{ph}\right>\approx1.08\times 10^3\si{m/s}$, and $\left<v_\mathrm{mag}\right>\approx0.87\times 10^4\si{m/s}$, which are of the same order of magnitude as the velocities evaluated based on the band calculations already presented in Fig.~\ref{fig:band}~(c).
Similarly, $\left<g_\mathrm{qp}\right>$ is qualitatively proportional to the inverse of bandwidth $W$.
Since $K_n$ contains $v^2g$, $\kappa$ is also in proportional to the square of $W$.
In the case of highly metallic materials such as Cu, $g_\mathrm{el}$ close to $\mu$ is significantly higher than for {BMB}. 
This is also true for semiconductors such as Si as well as many other insulators, as can be seen in Fig.~\ref{fig:k-materials}.
Consequently, these highly conductive metals generally exhibit very large $\kappa=500\si{W/mK}$ at room $T$.
Amongst the group-IV elements (C, Si and Ge), there is a monotonic decrease in the profile of $\kappa(T)$ as a consequence of the increasing elemental mass, which corresponds to a decrease in the strength of the lattice vibrational energy. 
However, in the case of Sn, $\kappa(T)$ instead continues to increase at low $T$s since the dominant QP changes from phonons to electrons, in line with the behavior of other metals such as Cu and Pt.

Compared to $v_\mathrm{qp}$ and $g_\mathrm{qp}$ which can be evaluated directly from band calculations, $\tau_\mathrm{qp}$ is a more challenging parameter to discuss and thus necessitates a more naive approach.
The parameter $\tau_\mathrm{el}$ is usually seen as being weakly $T$-independent, following from the Fermi statistics.
On the other hand, there is no such restriction upon the bosonic quasiparticles and their relaxation times depend largely upon $T$, generally decreasing as $T$ increases as a result of raising ph-ph and mag-mag scattering rates.
In insulators, this manifests as a sharp peak in $\kappa$ at low $T$s.
For instance in the case of high-purity diamond, one of the most thermally conductive materials, this peak reaches higher than $10^5 \si{W/mK}$~\cite{inyushkin_thermal_2018} at low $T$s after which point both ph-ph scattering and other extrinsic scatterings are exceeded by an increase in excitations via the $T$-dependence in the boson statistics.
Additionally, the phonon dispersion of diamond at around $100\si{meV}$ is much higher than other insulators and semiconductors, including BMB, resulting in its comparatively high thermal conductivity.

A general discussion for magnons is a little more complicated than for phonons, since the magnon dispersion at low energy is neither linear nor gapless.
The magnon dispersion of BMB is, in fact, parabolic due to its finite $18\si{meV}$ gap, resulting in magnon excitations being suppressed at low $T$s and a $\kappa_\mathrm{mag}$ which increases only slowly.
AFM insulators such as $\mathrm{Sr_2CuO_3}$, which exhibits magnon excitations with a large band-dispersion, as well as some other cuprates are also reported to have large $\kappa_\mathrm{mag}$~\cite{hess_heat_2019}.
These experimental data reflect the large magnon energy of around $300\si{\milli eV}$ for $\mathrm{La_2Cu_2O_3}$, which is greater than the phonon energy of diamond.
Similar profiles for additional AFM compounds can also be seen in Fig.~\ref{fig:k-materials}.
Even in these AFM compounds, however, the largest observation for $\kappa_\mathrm{mag}$ of c.a. 100 $\si{W/mK}$ is still smaller than the large $\kappa_\mathrm{ph}$ for diamonds, despite the large magnon velocity.
This difference can be understood in terms of the previously outlined scattering mechanism at the magnetic boundary, as well as additional relaxation paths existing in the case of magnetic insulators.
Regardless of the small value of $\kappa_\mathrm{mag}$, it is remarkable that the maximum of $\kappa_\mathrm{mag}$ can be observed at higher $T$s than that of $\kappa_\mathrm{ph}$, despite the existence of a finite gap and/or wider band dispersion.
It alludes to the possibility that excellent thermal conductors, determined by the contribution of $\kappa_\mathrm{mag}$, could be designed at room temperatures.
In addition, as a result of large anisotropy in the exchange interactions, a largely anisotropic $\kappa$ can be achieved via $\kappa_\mathrm{mag}$ without any special sample fabrication.
This can include interlayer rotated transition metal dichalcogenides, which show significant anisotropic $\kappa$ due to the breaking of interlayer translational symmetry~\cite{kim_extremely_2021}.

\section*{\label{sec:level3}Conclusion}
The longitudinal thermal-associated properties of the $\kappa$ and $S$ coefficients of BMB were studied, with and without longitudinal magnetic field, $H$.
In contrast to the small $H$ dependence of $\kappa$, the negative magneto-$S$ effect reaches a reduction of $38\%$ when $H$ is applied parallel to the $ab$-plane.
We have shown that numerical calculations, based on the BTE for electrons, phonons and magnons, can reasonably reproduce the experimental profiles of the $\kappa$ and $S$ coefficients.
An $H$-induced increase in $\kappa$ below $T\sim50\si{\kelvin}$ was observed as the AFM magnon excitation approaches the magnon energy gap, being indicative of the fact that the magnon-phonon interaction plays a significant role in determining the $\kappa$ of {BMB}.
Although the negative magneto-$S$ effect observed at low $T$s is consistent with the positive magneto-$\sigma$ previously reported in the same $T$ regime, the reason for the suppression of magneto-$S$ above $T\approx \SI{83}{K}$ (the temperature at which magneto-$\sigma$ almost vanishes) under an in-plane $H$ may remain an open question.
Increasingly sophisticated calculations, including interactions amongst electrons, phonons, and magnons, are anticipated to provide additional understanding in the future.
It is important to emphasize the fact that the contribution by magnons to $\kappa$ at relatively high $T$s can be comparable to that of phonons.
As such, utilizing magnons in AFM materials promises to be an excellent design methodology for applications requiring high performance thermal conductors at both intermediate and high $T$s.

\section*{\label{sec:level4}Methods}
Single crystals of BMB were synthesized via a self-flux method~\cite{saparov_crystals_2013}.
The crystal quality was then verified by X-ray diffraction (XRD) and energy-dispersive X-ray spectroscopy in a facility at the Advanced Institute of Materials Research, Tohoku University, Japan.
Considering the material softness, crystal XRD was measured along the crystallographic $c$-axis.
The field dependence of thermal conductivity $\kappa$ was measured using a custom-built measurement system operated with PPMS (Quantum Design) constructed at the Beijing Academy of Quantum Information Sciences (BAQIS), China.
For comparison with electrical conductivity $\sigma$, both heat current and magnetic field were applied along the $a$-axis.
The Seebeck coefficient, $S$, was measured using a custom-built measurement system at the High Field Laboratory for Superconducting Materials (HFLSM), Tohoku University. 
To eliminate the parasitic thermal voltage, the direction of the heat current was reversed for each measurement such that the data was collected in a reliable time scale.
The thermal conductivity $\kappa_\mathrm{qp}$ and Seebeck coefficient $S$ of the quasi particles (phonon, electron and magnon) were calculated within the framework of the Boltzmann transport equation, with additional calculations of $S$ via BoltzTraP2~\cite{madsen_boltztrap2_2018}. Details are provided in the Supplementary Information.

\section*{References}

%

\begin{acknowledgments}
Measurements of thermal transport properties were conducted at the High Field Laboratory for Superconducting Materials at Tohoku University, and the Nanoplat facilities at the Beijing Academy of Quantum Information Sciences.
This project has been supported by
National Natural Science Foundation of China (NSFC Grant Nos. 12174027,
11934001, 92265106, 11921005),
CREST Project by JST on Thermal Management,
Innovation Program for Quantum Science and Technology (2023ZD0300500),
National Key R\&D Program of China (Grant No. 2019YFA0308402),
Innovation Program for Quantum Science and Technology (Grant No. 2021ZD0302403),
Beijing Municipal Natural Science Foundation (Grant No. JQ20002).
J.-H.C acknowledges technical support form Peking Nanofab.
\end{acknowledgments}

\section*{\label{sec:level3}Appendix}
\appendix
\section{Calculation of thermal conductivity based on Boltzmann transport equation (BTE)}

In order to interpret the experimental data, the values of $\kappa_\mathrm{qp}$ (where qp $\in$ el, {ph}, and {mag}) have been calculated in the framework of the Boltzmann transport equation (BTE) under the relaxation time approximation with energy dependence.
In this context, the flux of the energy flow $J_\mathrm{qp}$ due to the temperature gradient $\bm{\nabla}_r T$ may be written as
\begin{align}
J_\mathrm{qp}(T) = \frac{1}{8\pi^3}\int_\mathrm{BZ}d\bm{k} n(\epsilon(\bm{k}),T)
	(\epsilon(\bm{k})-\mu) v(\epsilon(\bm{k})).
\end{align}
where only single band $\epsilon(\bm{k})$ is taken into account.
The non-equilibrium distribution function of $n(\epsilon(\bm{k}),T)$ in the presence of a uniform electric field $\bm{E}$ and $\bm{\nabla}_r T$ can be evaluated in the $\epsilon(\bm{k})$-dependent relaxation time approximation scheme as
\begin{align}
n(\bm{k},\epsilon,T)- f = \tau(\bm{k},\epsilon, T)\bm{v}(\bm{k})\left(-\frac{\partial f}{\partial \epsilon}\right)\left[-e\bm{\widetilde{E}}+\left(\frac{\epsilon - \mu(T)}{T}\right) (-\bm{\nabla}_r T) \right].
\end{align}
where $\epsilon(\bm{k})$, $\bm{v}(\bm{k}) \equiv \hbar^{-1}\nabla_k \epsilon$, $\tau(\bm{k},\epsilon,T)$, and $\mu(\bm{r},T)$ are the energy dispersion, group velocity, lifetime and chemical potential of a single QP, respectively, and $\bm{\widetilde{E}}=\bm{E}+\left(\frac{1}{e}\right)\bm{\nabla}_r \mu$ is the generalized electric field associated with $\mu$.
The equilibrium distribution function, $f$, represents either the Bose-Einstein $f_{BE}$ (for phonons and magnons) or Fermi-Dirac $f_{FD}$ (for electrons) statistics.
Finally, the operator $\bm{\nabla}_r$ denotes the first-order differential in the real space, and the integral spans the entire first Brillouin zone (BZ) in $k$-space.

Consequently, the flow of thermal energy along the longitudinal direction associated with each QP is described by the expression:
\begin{align}
J_\mathrm{qp}=\frac{1}{8\pi^3}\int_\mathrm{BZ}d\bm{k}\tau v^2  \left(-\frac{\partial f}{\partial \epsilon}\right) (\epsilon-\mu) \left[(-e)\bm{\widetilde{E}}+\frac{(\epsilon-\mu)}{T}(-\bm{\nabla}_r T)\right].
\end{align}
We take the directions of both the quasiparticle flux and the thermal gradient to lie along the longitudinal $a$-axis of BMB and henceforth omit the directional index.
Following from the definition of thermal current $J_\mathrm{qp}=\kappa_\mathrm{qp}\bm{\nabla}_r T$~\cite{grosso_stp_ch11}, in the case of electrons where contributions to $\bm{\widetilde{E}}$ from both $\bm{E}$ and $\bm{\nabla}_r \mu$ should be taken into account, $\kappa_\mathrm{el}(T)$ can written in terms of the kinetic coefficient 
\begin{align}
	K_{n}\equiv \frac{1}{8\pi^3}\int_\mathrm{BZ} d\bm{k} \tau v^2(\epsilon-\mu)^n\left(-\frac{\partial f}{\partial \epsilon}\right) 
\end{align}
with $f=f_{FD}$ the Fermi-Dirac distribution function, as 
\begin{align}
	\kappa_\mathrm{el}(T)=\frac{1}{T}\left(K_2-\frac{K_1^2}{K_0}\right).
\end{align}
On the other hand, for bosons without elementary charge such as phonons and magnons, the $\mu$ part can be safely ignored for longitudinal thermal transport and therefore $\bm{\widetilde{E}}=0$.
The T-dependent thermal conductivity in the longitudinal direction $\kappa_\mathrm{qp}(T)$ for bosonic quasiparticles can thus be written as
\begin{align}
\kappa_\mathrm{qp} (T)   = \frac{1}{8\pi^3}\int_\mathrm{BZ}d\bm{k}(1/T) \tau v^2 (\epsilon-\mu)^2 \left(-\frac{\partial f}{\partial \epsilon}\right).
\end{align}
This can also be written in terms of the $K_n$ with $f_{BE}$ statistics as $\kappa_\mathrm{ph,mag} (T)=K_2/T$. 

\section{Calculation of dispersion relation and transport properties}
The electronic band dispersion relation of BMB was calculated via spin-polarized band calculations using the WIEN2K~\cite{wien-2K} and Quantum ESPRESSO~\cite{QE-2009,QE-2017} packages. The lattice parameters $a \, = \, b$, $c$ and $z_{Bi}$ are adapted from the previous study~\cite{saparov_crystals_2013} and the generalized gradient approximation (GGA) by Perdew, Burke, and Ernzerhof (PBE)~\cite{perdew_generalized_1996} exchange-correlation potential was chosen. The results of calculation via these two software don't show any significant differences.
The phonon dispersion is obtained from first principle calculations using VASP~\cite{kresse_ab_1993,kresse_ab_1994,kresse_efficiency_1996,kresse_efficient_1996} and open source code Phonopy~\cite{phonopy}.
For the magnon dispersion, an analytical form of an XXZ Heisenberg model is adopted from the previous report~\cite{calder_magnetic_2014} for explaining the neutron diffraction experimental data, written as,
\begin{align}
	\mathcal{H}_{XXZ} = \sum_{i,j}J_{i,j}\vec{S}_i\cdot\vec{S}_j + \sum_i D{(S^z_i)}^2,
	\label{eq:Heisenberg}
\end{align}
\noindent
where $J_{i,j}$ is the exchange interaction between two local moments located at sites $i$ and $j$, and $D$ is an anisotropic energy that aligns magnetic moments along the $c$-axis as in the magnetization experiment results.
The nearest neighbour in the $ab$-plane, second-nearest neighbour in the $ab$-plane, and nearest neighbour in the $c$-axis are taken into account.
Spin operators $\vec{S}_i$ and $S^z_i$ in $\mathcal{H}_{XXZ}$ are converted to bosonic creation and annihilation operators via the Holstein Primakoff transformation, and $\mathcal{H}_{XXZ}$ is then diagonalized by the Bogoliubov transformation~\cite{rezende_introduction_2019}.
Due to our interest in the in-plane transport phenomena, we calculated the $x$ component of the group velocity $v_k = \hbar^{-1}{(\bm{\nabla}_k \epsilon)}_x$ of electrons, phonons and magnons.
We calculated the derivatives along the path between symmetric k-points in the $k_x-k_y$ plane such as $M-\Gamma$ in Fig.~\ref{fig:band}.
The density of states ($g$) and velocity ($v$) spectra are then calculated from the dispersion relations (Fig.~\ref{fig:band}~(b, c)).

The thermal conductivity $\kappa$ and Seebeck coefficient $S$ associated with the electrons are calculated via BoltzTraP2~\cite{madsen_boltztrap2_2018}, a program calculating semi-classical transport coefficients based on the BTE and the band structure results obtained from DFT calculations~\cite{ogasawara_magnetic_2022} in a framework of constant relaxation time ($\tau = const.$).
The relaxation time of electrons was chosen by comparing the theoretical evaluation and experimental investigations of the electrical conductivity $\sigma$ that had been reported previously~\cite{ogasawara_large_2021}.
The temperature dependence of the chemical potential was determined from the conservation of carrier number, whereby we calculate $n(\mu) = \int d\epsilon g(\epsilon) f_{FD}(\epsilon,\mu)$ at each temperature and find $\mu$ such that $n(\mu)$ aligns with the experimental results.
Experimental data from previous Hall measurements were used as the reference for the carrier number~\cite{ogasawara_large_2021}.

\section{Calculation of relaxation time}
In the case of phonon and magnon, $\tau$ cannot be estimate from the electric conductivity as well as electron due to no contirbution.
In order to calculate $\kappa$ of phonon and magnon, the relaxation time generally requires further parametrization.
Considering the prohibitively large scale of the calculations required for a complete treatment of the scattering cross sections, the calculation of the relaxation rate of intrinsic scattering processes such as phonon-phonon scattering are frequently approximated by employing empirical formulae.
For phonons, the contribution of three-phonon scattering (including Umklapp scattering) and boundary scattering were taken into account as follows~\cite{walker_phonon_1963},
\begin{align}
	\tau_\mathrm{ph} \approx \eta^{-1}_\mathrm{ph} & = {(\eta_\mathrm{boundary} + \eta_\mathrm{ph-ph})}^{-1}, \\
	\eta_\mathrm{boundary}                         & = v_\mathrm{ph}/L_l,                                     \\
	\eta_\mathrm{ph-ph}                            & = a \epsilon^2 T \exp(-b/T),
	\label{eq:ph-ph}
\end{align}
\noindent
with $L_l$ being the characteristic distance between boundaries. There are three free parameters, $\eta_\mathrm{boundary}$, $a$ and $b$.
To reduce the number of fitting parameters, we choose to omit the remaining terms included in Ref.~\cite{walker_phonon_1963}.
In the case of magnons, contributions from 4-magnon scattering and magnetic boundary scattering were considered~\cite{rezende_magnon_2014,bayrakci_lifetimes_2013},
\begin{align}
	\tau_\mathrm{mag}      & \approx \eta^{-1}_\mathrm{mag} = {(\eta_\mathrm{boundary} + \eta_\mathrm{mag-mag})}^{-1}, \\
	\eta_\mathrm{boundary} & = v_\mathrm{mag}/L_m,                                                                     \\
	\eta_\mathrm{mag-mag}  & = cq^2T^2 , q = k/k_\mathrm{max}.
	\label{eq:mag-mag}
\end{align}
Since second order terms of the post-diagonalization Heisenberg Hamiltonian are fourth order in the annihilation/creation operators, three-magnon scattering is ignored here.
We must note here that the $k$ and $T$ dependence of $\eta_\mathrm{mag-mag}$ is cited from previous research for YIG that is not AFM, but FM (or, more precisely, ferrimagnetic).
Despite this distinction, we assume a similar dependence for $\eta_\mathrm{mag-mag}$ since both materials exhibit a quasi-parabolic magnon dispersion near the BZ center from which the dominant contribution arises.

\noindent
Using the $K_n$ we can write $\sigma$, $S$, and $\kappa_\mathrm{el}$ in compact form as $\sigma=e^2K_0$, $S=-\frac{1}{eT}\frac{K_1}{K_0}$, and $\kappa_\mathrm{el}=\frac{1}{T}\left(K_2-\frac{K_1^2}{K_0}\right)$, respectively~\cite{grosso_stp_ch11}.
Since the second term of $\kappa$ is related to the electronic contributions, we can ignore it in the case of phonons and magnons.
We further note that although the above equations are written here for one band only, in reality we generally consider all bands and their degeneracies.

\section{Calculation of specific heat}
The Boltzmann transport equation used herein contains an inexactness due to our approximation of the relaxation time, $\tau$. On the other hand, static properties such as specific heat are independent of $\tau$ and we therefore need only employ the dispersion relation to calculate these quantities.
Here, we calculate the specific heat of the phonon and magnon to confirm the validity of the calculations and dispersion relation, $\epsilon$, used in the main text.
Ignoring the $T$-dependence of $\epsilon$, the specific heat may be straightforwardly calculated via,
\begin{align}
	C_P \approx C_V = \frac{\partial E}{\partial T} = \frac{1}{8\pi^3}\int_\mathrm{BZ} d\bm{k}\epsilon \frac{\partial f_{BE}}{\partial T},
\end{align}
where $f_{BE}$ is the Bose-Einstein distribution function. The result of this calculation is shown in Fig.~\ref{fig:cp}, alongside previous experimental results~\cite{ogasawara_large_2021}. We show as horizontal lines the Dulong-Petit law, the saturated specific heat described by the number of degree of freedom in a unit cell $N$ with molar gas constant $R$, i.e. $C_P \approx NR[\si{J/K/\mole}]$. Similar to $\kappa$, the phonon profile alone reproduces the low-$T$ region, with the magnon contribution becoming appreciable around $T\sim50\si{K}$. While there exists small deviation from the experimental data, possibly arising from the $T$ dependence of the phonon (and magnon) dispersions, it shows a good degree of agreement. We note that the electron's contribution to $C_P$, much like $\kappa$, is likewise negligibly small.

\section{Calculation of magnon dispersion under magnetic field}
The magnon dispersion is derived from the Heisenberg spin Hamiltonian,
\begin{align}
	\mathcal{H} & = \sum_{i,j}J_{i,j}\bm{S}_i\cdot\bm{S}_j + \sum_iD(S^z_i)^2 \nonumber                                                       \\
	            & = \sum_{i,j}J_{i,j}\left\{S^z_iS^z_j + \frac{1}{2}\left(S^+_iS^-_j + S^-_iS^+_j\right)\right\} + \sum_iD(S^z_i)^2  \\
	S^\pm_i     & = S^x_i \pm iS^y_i \nonumber
\end{align}
The first and second terms give the exchange interaction and anisotropy, respectively. Because the spin operator has an upper limit, it cannot be treated the same as the quasiparticle creation/annihilation operators. Application of the Holstein-Primakoff transfomation~\cite{holstein_field_1940} resolves this problem,
\begin{align}
	S_i^z & \rightarrow S-a^\dagger_ia_i, \nonumber                                             \\
	S_i^+ & \rightarrow (2S)^{1/2}\left(1-\frac{a^\dagger_ia_i}{2S}\right)^{1/2}a_i,        \\
	S_i^- & \rightarrow (2S)^{1/2}a^\dagger_i\left(1-\frac{a^\dagger_ia_i}{2S}\right)^{1/2} \nonumber
\end{align}
As the transformation of $S_i^z$ indicates, the creation/annihilation operator gives the number of spins deviating from a completely ordered state. In the case of AFM, it is noted that the creation/annihilation operators should be defined in each sublattice omitted here. When the number of excitations $a^\dagger_ia_i$ is much smaller than $2S$, the square root in $S_i^\pm$ can be expanded, resulting sort of even order terms against creation/annihilation operators. Following this, the Fourier and Bogoliubov transformations diagonalize the second order terms of the spin Hamiltonian, with higher-order terms indicating the magnon-magnon scattering.

When the transverse magnetic field is applied, Zeeman terms such as $g_mu_BH\sum_iS^x_i$ are introduced into the spin Hamiltonian. Here, the magnetic field is assumed to be applied along $x$ direction. Hence, $S^x_i$ is expanded to odd order terms in the creation/annihilation operators, the Zeeman term cannot be implemented into the diagonalization process mentioned above. Alternatively, we consider the spin canting from the zero field state. The canting angle can be estimated classically by minimizing the energy of the Heisenberg Hamiltonian with the Zeeman term included. The canting results in a rotation of the principle axis of spin. In the case of ferromagnets, it does not produce any change in the exchange term because their relative angles are fixed. On the other hand, AFM features sublattices and relative angle changes. The magnon dispersion of a canted AFM state is obtained by diagonalization of a new spin Hamiltonian. This methodology successfully reproduces the experimental data of MnF$_2$~\cite{rezende_introduction_2019}. On the other hand, in the case of lontgitudinal field, the Zeeman term $g_mu_BH\sum_iS^z_i$ is commutative with Heisenberg hamiltonian. Therefore, it is simply introduced as Zeeman term.
\setcounter{figure}{0}
\renewcommand{\figurename}{SupplFig.}
\begin{figure}[ht]
	\centering
	\includegraphics[width=0.49\textwidth]{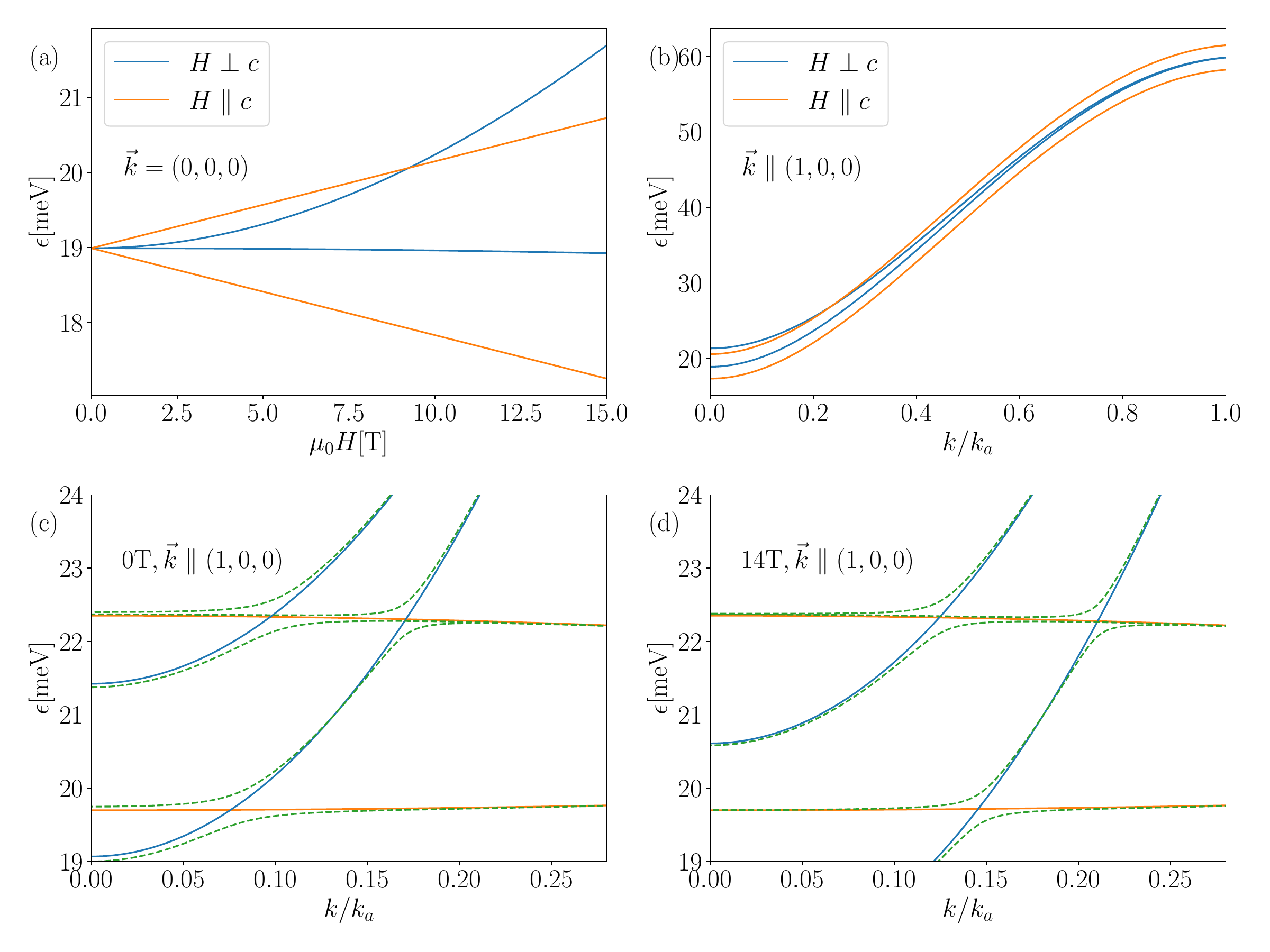}
	\caption{(a) Magnetic field dependence of the magnon dispersion at the $\Gamma$-point. Twofold degeneracy is lifted off at finite magnetic field $H$. (b) The $k$-dependence of the magnon dispersion at $\mu_0H=\SI{14}{T}$ (orange and blue lines). Dotted lines indicate the dispersion in the absence of a magnetic field. (c,d) Magnon (blue) and phonon (orange) dispersion focusing on the bottom of magnon dispersion with a $\SI{14}{T}$ magnetic field (c) $H\perp c$ and (d) $H\parallel c$ configuration. Green curves indicate the magnon-phonon coupling state assuming a coupling constant $g=J_1\times 10^-2=0.127\si{meV}$. }\label{fig:mag-ph}
\end{figure}

\section{Magnon-phonon coupling state}
Theoretically, magnons and phonons are coupled via a magneto-elastic effect that is described by the Hamiltonian $\mathcal{H}_\mathrm{ME}$ as follows~\cite{simensen_magnon-polarons_2019},
\begin{align}
	\mathcal{H}_\mathrm{ME} = \sum_{\alpha, \beta}\sum_{i,\delta}B^{\alpha \beta}S_i^\alpha S_{i+\delta}^\beta R^{\alpha \beta}_{i,i+\delta}
\end{align}
where $\alpha$ and $\beta$ are directional indices of spins $S^{\alpha / \beta}_i$ and local strain $R^{\alpha \beta}_{i,i+\delta}$. $B^{\alpha \beta}$ is coupling constant that is described as $B^{\alpha \beta}=\delta^{\alpha \beta}B^\parallel + (1-\delta^{\alpha \beta})B^\perp$. In the case of a transverse mode ($\alpha \neq \beta$) involving the $z$-direction, e.g. $(\alpha, \beta) = (x,z)$, these terms contain $S^x$ or $S^y$ and $R^{\alpha \beta}_{i,i+\delta}$. They are transformed to first order in the creation/annihilation operators which couple the magnons and phonons. If we assume creation/annihilation operators of the magnons and phonons to be $a/a^\dagger$ and $c/c^\dagger$, respectively (omitting indices for simplicity), the simplest form the Hamiltonian can take is,
\begin{align}
	\mathcal{H}_\mathrm{mg-ph} = \hbar\omega a^\dagger a + \hbar\Omega c^\dagger c + g(a^\dagger c + c^\dagger a).
\end{align}
Here, $\omega$ and $\Omega$ are the eigen-frequencies of the magnon and phonon, and $g$ represents the strength of coupling, determined from $B^{\alpha \beta}$. After diagonalization, the magnon-phonon hybridized states are obtained as shown in the main text.


\setcounter{figure}{1}
\renewcommand{\figurename}{SupplFig.}
\begin{figure}[ht]
	\centering
	\includegraphics[width=0.48\textwidth]{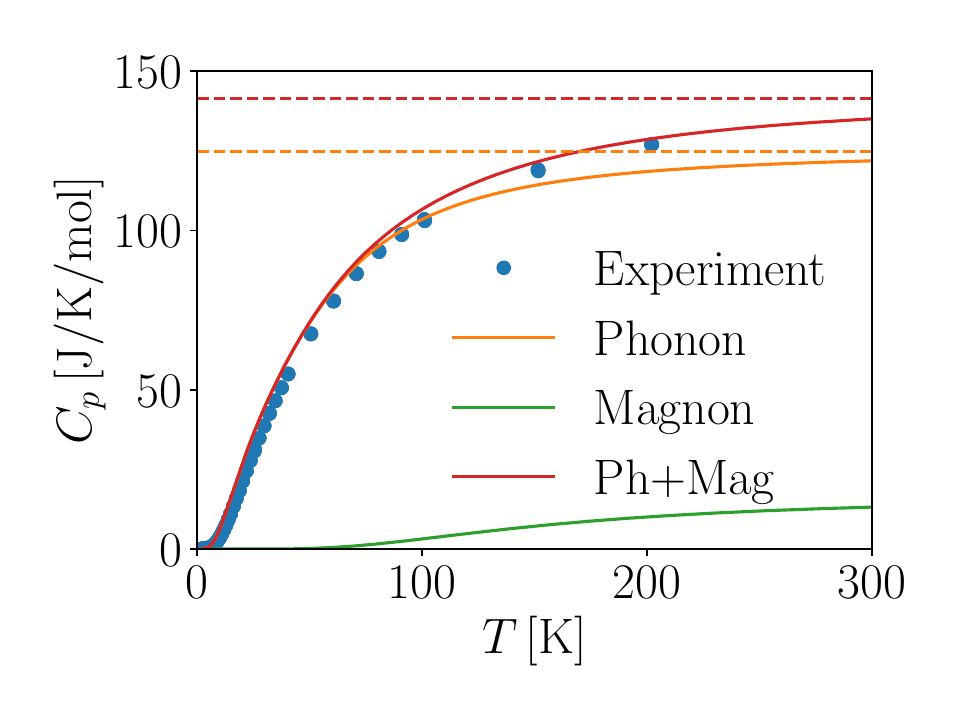}
	\caption{Specific heat of the phonon and magnon as compared to experimental results. Dashed lines give the value of the Dulong-Petit law for which only the phonon (orange) and phonon + magnon (red) contributions are taken into account.}\label{fig:cp}
\end{figure}



\end{document}